\documentclass[aps,prl,showkeys,showpacs,reprint]{revtex4-1}
	
	\usepackage{setspace}
	\usepackage{amsmath}    
	\usepackage{graphicx}   
	\usepackage{verbatim}   
	\usepackage{color}      
	\usepackage{subfigure}  
	\usepackage{hyperref}   
		
\makeindex

\begin{document}


\pagestyle{myheadings}

\title{High-precision spectroscopy of the forbidden $2 \ ^3\text{S}_1 \to 2 \ ^1\text{P}_1$ transition in quantum degenerate metastable helium}

\author{R.P.M.J.W. Notermans and W. Vassen}
\email{w.vassen@vu.nl}
\affiliation{LaserLaB, Department of Physics and Astronomy, VU University Amsterdam \\ De Boelelaan 1081, 1081 HV Amsterdam, The Netherlands}


\begin{abstract}
We have measured the forbidden $2 \ ^3\text{S}_1 \to 2 \ ^1\text{P}_1$ transition at 887~nm in a quantum degenerate gas of metastable $^4$He atoms confined in an optical dipole trap. The determined transition frequency is 338\,133\,594.4~(0.5)~MHz, from which we obtain an ionization energy of the $2 \ ^1\text{P}_1$ state of 814\,709\,148.6~(0.5)~MHz. This ionization energy is in disagreement by $> 3\sigma$ with the most accurate quantum electrodynamics (QED) calculations available. Our measurements also provide a new determination of the lifetime of the $2 \ ^1\text{P}_1$ state of $0.551 \ (0.004)_{\text{stat}} \ (^{+ 0.013}_{- 0.000})_{\text{syst}}$~ns, which is the most accurate determination to date and in excellent agreement with theory.
\end{abstract}


\maketitle

Quantum electrodynamics (QED) is one of the most thoroughly tested theories in physics. From QED theory and accurate measurements, the fine structure constant \cite{Aoyama1, Hanneke1}, the Rydberg constant \cite{Biraben1}, nuclear charge radii \cite{Antognini1,Rooij} and the electron mass can be deduced \cite{Sturm1}. It can also provide accurate ionization energies for one- and two-electron atoms. To test QED, both highly accurate calculations and high-precision experimental data are required. Few-body systems such as the hydrogen atom and helium atom are candidates that fulfill both criteria. Testing and applying QED in these systems has led to surprising results in recent years. An example is the $7\sigma$ discrepancy in the proton size derived from muonic hydrogen Lamb shift measurements and the accepted CODATA value, also referred to as the proton size puzzle \cite{Antognini1,ProtonPuzzle}. Recent measurements of the helium $2 \ ^3\text{S} \to 2 \ ^1\text{S}$ transition at 1557~nm \cite{Rooij} and the $2 \ ^3\text{S} \to 2 \ ^3\text{P}$ transitions at 1083~nm \cite{CancioPastor1} disagree by $4\sigma$ in the determination of the helium isotopic nuclear size difference. These measurements provide a unique comparison with nuclear size measurements in the muonic helium ion, developed to help solve the proton size puzzle \cite{Nebel1}.

\begin{figure}[tbp]
	\begin{center}
	\includegraphics[width=0.4\textwidth]{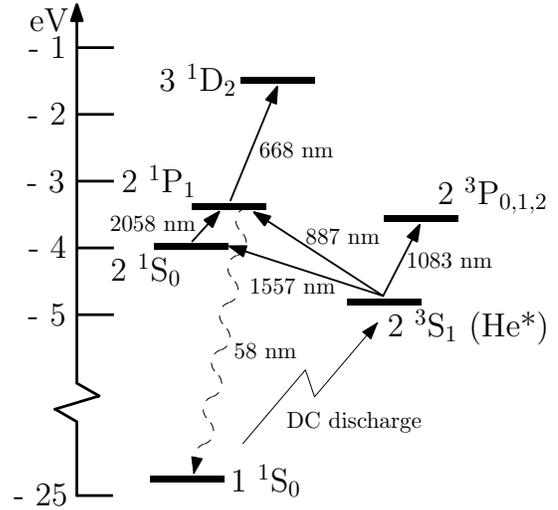}
	\caption{Schematic overview of the energies of the lowest levels in $^4$He (with respect to the ionization limit) and the transitions that are mentioned in this work. The $2 \ ^3\text{S}_1 \to 2 \ ^3\text{P}_{2}$ transition at 1083~nm is used for laser cooling and trapping.}
	\label{fig:HeLevelScheme}
	\end{center}
\end{figure}

In particular for the low-lying states with low angular momentum, accurate measurements of the ionization energies (IE) in helium have allowed stringent tests of two-electron QED \cite{Rooij,CancioPastor1,Lichten1,Sansonetti2,Dorrer1,CancioPastor2,Kandula1}. A schematic overview of the lowest states of helium together with transition wavelengths mentioned in this paper are shown in Figure~\ref{fig:HeLevelScheme}. In comparing the experimentally determined IE to QED calculations, a discrepancy of 6.5~(3.0)~MHz in the $2 \ ^1\text{P}_1$ IE was identified by Drake and Pachucki \cite{Pachucki1,*Pachucki1_erratum,Drake1,Pachucki2}. This discrepancy is based on a measurement of the $2 \ ^1\text{P}_1 \to 3 \ ^1\text{D}_2$ transition frequency with 3~MHz accuracy by Sansonetti and Martin in 1984 \cite{Sansonetti1}. As the QED calculation of this IE is accurate to $0.4$~MHz \cite{Pachucki2}, a more accurate measurement should be able to determine whether this discrepancy still stands. Recently, two new determinations of the $2 \ ^1\text{P}_1$ IE were reported by Luo \emph{et al.} based on the measurements of the $2 \ ^1\text{S}_0 \to 2 \ ^1\text{P}_1$ \cite{Luo1,*Luo1_erratum} and $2 \ ^1\text{P}_1 \to 3 \ ^1\text{D}_2$ \cite{Luo2} transition frequencies. As these transitions are electric dipole-allowed, the measurements could be done using saturated absorption spectroscopy in an RF discharge cell. The extracted ionization energies for the $2 \ ^1\text{P}_1$ state disagree with QED theory at the $3.5\sigma$ level.

In this work we report the direct measurement of the forbidden $2 \ ^3\text{S}_1 \to 2 \ ^1\text{P}_1$ transition at 887~nm in a quantum degenerate gas (QDG) of metastable $2 \ ^3\text{S}_1$ state helium ($^4$He*, lifetime $\approx$ 7800 s) atoms confined in an optical dipole trap (ODT). The advantage of performing spectroscopy in an ODT is the ability to probe very weak transitions and the simultaneous reduction and characterization of systematic effects to the kHz level \cite{Rooij}. As the theoretical natural linewidth of this transition is 287~MHz \cite{Morton2}, the accuracy of our measurement is limited by statistics rather than by systematic effects. Combined with the accurately known IE of the $2 \ ^3\text{S}_1$ state this measurement of the transition frequency enables a determination of the $2 \ ^1\text{P}_1$ IE.

\begin{figure}[tbp]
	\begin{center}
	\includegraphics[width=0.4\textwidth]{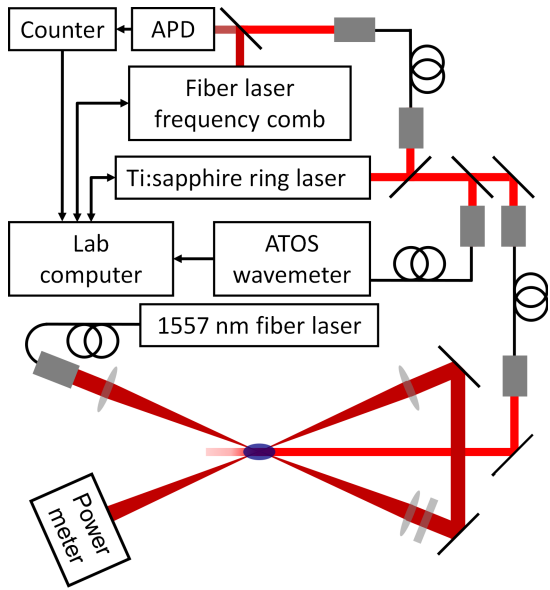}
	\caption{(Color online) Schematic overview of the setup. The crossed dipole trap is created using a fiber laser and the trapping beam power is measured using a power meter. The spectroscopy light is generated using a Ti:sapphire ring laser. The beatnote between the spectroscopy laser and the frequency comb is measured with an avalanche photodiode (APD) connected to a frequency counter and digitally sent to the lab computer. The computer then calculates and sends a proportional-integral (PI) feedback signal to the Ti:sapphire laser to stabilize the spectroscopy laser frequency. The lab computer also interfaces with the frequency comb, enabling us to register and control the frequency comb settings.}
	\label{fig:setup}
	\end{center}
\end{figure}

The measured lineshape of the transition allows for an accurate determination of the lifetime of the $2 \ ^1\text{P}_1$ state. This method does not require the branching ratios of decay channels, which is the main problem in fluorescence measurements of the lifetime \cite{Fry1,*Martinson1,*Burger1,*Larsson1,*Zitnik1}, and the only dominant broadening effect in our experiment can be calculated using the optical Bloch equations.

The $2 \ ^3\text{S}_1 \to 2 \ ^1\text{P}_1$ transition is forbidden as it violates conservation of spin. Due to a small mixing of the $2 \ ^1\text{P}_1$ and $2 \ ^3\text{P}_1$ states \cite{Drake3} the electric dipole transition has an Einstein A coefficient of $1.4423 \ \text{s}^{-1}$, which is seven orders of magnitude weaker than regular dipole-allowed transitions in the helium atom \cite{Drake4}. Therefore this transition has, to our knowledge, never been observed before. In order to obtain a good signal with reasonable laser power, the atoms need to be probed on a timescale of about $1$ s, and we achieve this by trapping a QDG of $^4$He* atoms in an ODT. For this we use the same experimental setup as used to measure the doubly forbidden $2 \ ^3\text{S} \to 2 \ ^1\text{S}$ transition \cite{Rooij}. We produce a QDG consisting of a thermal gas and a Bose-Einstein condensate (BEC) in a crossed-beam ODT, which is created using an NP Photonics fiber laser operating at a wavelength of 1557.3~nm. Details on the production and physics of ultracold metastable gases can be found in \cite{Vassen1}. The ODT is kept shallow at a depth of about $1.3 \ \mu\text{K}$ to minimize systematic shifts. After thermalization the temperature of the gas is approximately $0.2 \ \mu\text{K}$. We apply a small homogeneous magnetic field in the ODT to maintain spin polarization of the gas, which is required to have a trap lifetime $> 10$~s. The small Zeeman shift is directly measured using RF transitions between the $2 \ ^3\text{S}_1 \ M_J = +1,0,-1$ states with kHz accuracy and therefore does not provide a limitation for our experimental accuracy \cite{Rooij}.

Once the QDG is loaded in the ODT, an approximately $50$~mW probe beam excites the atoms to the $2 \ ^1\text{P}_1$ state during 1~s. The excited atoms decay in 0.5~ns to the $1 \ ^1\text{S}_0$ state and leave the trap. Then the ODT is turned off and the remaining atoms fall due to gravity and hit a microchannel-plate (MCP) detector. The MCP current is measured to determine the time-of-flight (TOF) distribution of the atoms. This TOF distribution is fit using a bimodal distribution, which describes the momentum distribution of the BEC and thermal fractions. From the fit we obtain the atom number of both fractions, the temperature of the thermal fraction and the chemical potential of the BEC \cite{Rooij}.

\begin{figure}[tbp]
	\begin{center}
		\includegraphics[width=0.4\textwidth]{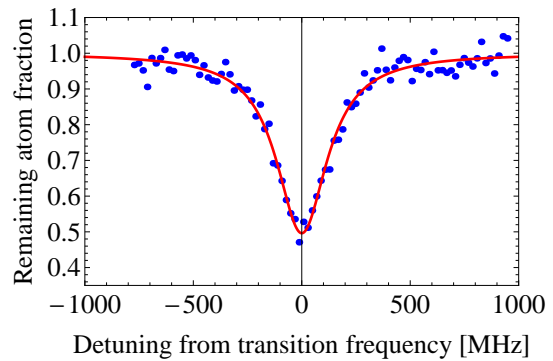}
		\caption{(Color online) Example of a single linescan measurement, showing 90 consecutive measurements over a range of approximately $ 1.8$~GHz. A Lorentzian fit is used to determine the transition frequency and the linewidth. The frequency axis is centered on the transition frequency determined in this scan.}
		\label{fig:LineScan_example}
	\end{center}
\end{figure}

A schematic overview of the ODT and the metrology infrastructure is shown in Figure~\ref{fig:setup}. The probe beam is generated using a Coherent 899-21 Ti:sapphire laser with an output power of 0.4~W at 887~nm. During the measurements the wavelength is registered using a wavemeter. Simultaneously we use an erbium-doped fiber laser frequency comb that is stabilized using a GPS-controlled rubidium clock to create a beatnote with the probe laser \cite{Rooij}. Combining the wavemeter data with the beatnote data provides the absolute frequency of the probe laser.

Additionally, we stabilize the Ti:sapphire laser frequency to the frequency comb using a proportional-integral (PI) control loop. We control the Ti:sapphire laser frequency by keeping the beatnote frequency constant and scanning the repetition rate of the frequency comb. Due to the relatively slow loop time of 30~ms of the PI control loop, our laser has a Gaussian lineshape with an average FWHM of approximately $1$~MHz, with an accuracy of $<5$~kHz, during the measurements.

\begin{figure}[tbp]
	\begin{center}
		\includegraphics[width=0.4\textwidth]{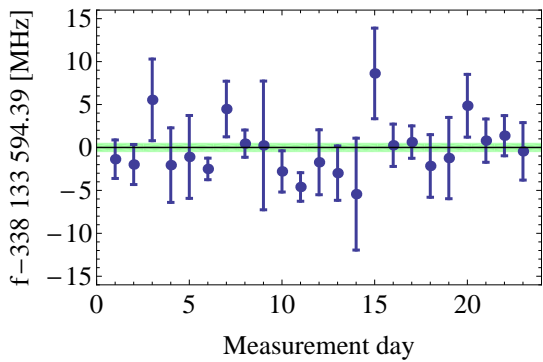}
		\caption{(Color online) Determined transition frequency averaged per measurement day, based on a total of 77 measurements. The error bars on the data represent the $1\sigma$ standard deviation of the daily average. The frequencies are centered around the final average transition frequency and the green bar represents its $1\sigma$ standard deviation of 0.5~MHz.}
		\label{fig:Daily_Averages}
	\end{center}
\end{figure}

In our experiment we measure a linescan over the resonance by 90 individual measurements with a frequency stepsize of 20~MHz to obtain a normalized loss profile as seen in Figure~\ref{fig:LineScan_example}. A fit with a Lorentzian lineshape function is used to obtain the linewidth and the central frequency \cite{OnlineMaterial}.

The advantage of doing spectroscopy of a QDG in an ODT is the high degree of control over systematic effects. Ab-initio calculations of the polarizabilities of both the $2 \ ^3\text{S}_1$ and $2 \ ^1\text{P}_1$ states at both the wavelength of the ODT and the spectroscopy beam are combined with previously performed AC Stark shift measurements \cite{Rooij}. The resulting AC Stark shift of the measured transition is 31~kHz, calculated with kHz precision. The Zeeman shift is measured to kHz precision as well. The recoil shift is 63.5~kHz, calculated with sub-kHz accuracy. Broadening effects due to the finite size of the QDG in the ODT and due to the momentum distribution of the gas \cite{Killian1} are below 50 kHz and therefore negligible as well. The mean field shift cannot be calculated directly as the $2 \ ^3\text{S}_1 - 2 \ ^1\text{P}_1$ cold-collision scattering length is not known. However, the range of possible mean field shifts can be calculated based on a model of Kokkelmans \emph{et al.} \cite{Kokkelmans1} and the known $^5 \Sigma^+_g$ $2 \ ^3\text{S}_1 - 2 \ ^3\text{S}_1$ scattering length of 142.0~(1)~$a_0$, where $a_0$ is the Bohr radius \cite{Moal1}. From this model we find a worst-case mean field shift of 90 kHz at extraordinary large scattering lengths \cite{OnlineMaterial}. However, the actual mean field shift is expected to be much smaller as the finite lifetime of the $2 \ ^1\text{P}_1$ state reduces the mean field interaction \cite{Julienne1}.

Based on a total of 77 linescans taken over a period of two months in summer 2013, the daily average transition frequency is shown in Figure~\ref{fig:Daily_Averages}. We obtain a $2 \ ^3\text{S}_1 \to 2 \ ^1\text{P}_1$ transition frequency of 338\,133\,594.4~(0.5)~MHz. This value is in good agreement with the most recent theoretical value of 338\,133\,594.9~(2.7)~MHz \cite{Pachucki2}, where the accuracy is limited by the QED calculations of the $2 \ ^3\text{S}_1$ state.

\begin{figure}[tbp]
	\begin{center}
		\includegraphics[width=0.4\textwidth]{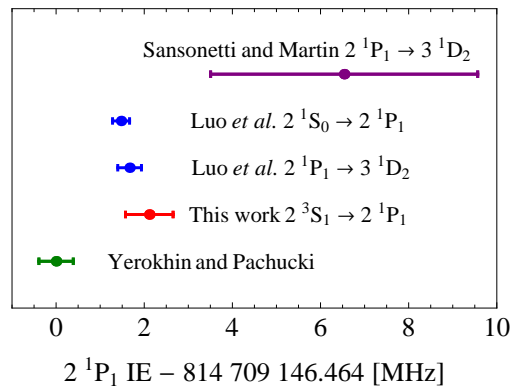}
		\caption{(Color online) Comparison of our experimental result for the $2 \ ^1\text{P}_1$ IE with other experiments \cite{Luo1,*Luo1_erratum,Luo2,Sansonetti1} and QED theory by Yerokhin and Pachucki \cite{Pachucki2}. All recent experimental results show a $> 3\sigma$ discrepancy with theory.}
		\label{fig:Results_All}
	\end{center}
\end{figure}

From our previously measured $2 \ ^3\text{S}_1 \to 2 \ ^1\text{S}_0$ transition frequency (192\,510\,702.1456~(0.0018)~MHz \cite{Rooij}) we extract a $2 \ ^1\text{S}_0 \to 2 \ ^1\text{P}_1$ transition frequency of 145\,622\,892.2~(0.5)~MHz. This result agrees with the recent  $2 \ ^1\text{S}_0 \to 2 \ ^1\text{P}_1$ frequency measurement \cite{Luo1,*Luo1_erratum} within 0.6~(0.6)~MHz. The $2 \ ^3\text{S}_1$ IE of 1\,152\,842\,742.97~(0.06)~MHz, derived from a measurement of the $2 \ ^3\text{S}_1 \to 3 \ ^3\text{D}_1$ transition frequency \cite{Dorrer1} and the calculated $3 \ ^3\text{D}_1$ IE \cite{Drake2}, can now be combined with our result to give a $2 \ ^1\text{P}_1$ IE of 814\,709\,148.6~(0.5)~MHz. Comparing this result to both measurements of Luo \emph{et al.} \cite{Luo1,*Luo1_erratum,Luo2}, we find very good agreement. An overview of the most accurate experimental results and the QED calculations for the $2 \ ^1\text{P}_1$ IE is shown in Figure~\ref{fig:Results_All}. A discrepancy of $>3\sigma$ with the theoretical IE as calculated by Yerokhin and Pachucki \cite{Pachucki2} remains. As QED calculations of most low-lying states of $^4$He agree very well with experiment, improved calculations for the $2 \ ^1\text{P}_1$ state are now timely. It may be that the contribution of $m \alpha^7$ terms is not treated well in this case, as a re-evaluation of these terms has shifted the IE by almost 1~MHz \cite{Pachucki1,*Pachucki1_erratum,Pachucki2}.

The fit of our measurements to a Lorentzian line profile also provides the linewidth of the $2 \ ^1\text{P}_1$ state, which is among the largest of all atomic transitions in Nature. Due to the finite number of atoms in the trap, the lineshape of the trap loss signal is broadened as, even for an off-resonance laser beam, the number of atoms left in the trap will be zero for infinite interaction time. Therefore we calculate the population of the $2 \ ^3\text{S}_1$ state using a three-level optical Bloch equations model based on the model used by Van Leeuwen and Vassen \cite{Leeuwen1}, and use it to correct for this broadening effect \cite{OnlineMaterial}.

Saturation of the MCP detector can lead to a broadening effect at the 1 MHz level of accuracy at which we determine the linewidth. Although saturation effects are expected and have been observed in other experiments with metastable helium BECs \cite{Schellekens1}, analysis of our data shows no statistical significant broadening due to saturation. A systematic uncertainty is added to the result to indicate the worst-case shift in the linewidth if we allow a nonlinear response of the MCP detector in our analysis~\cite{OnlineMaterial}.

\begin{figure}[tbp]
	\begin{center}
		\includegraphics[width=0.4\textwidth]{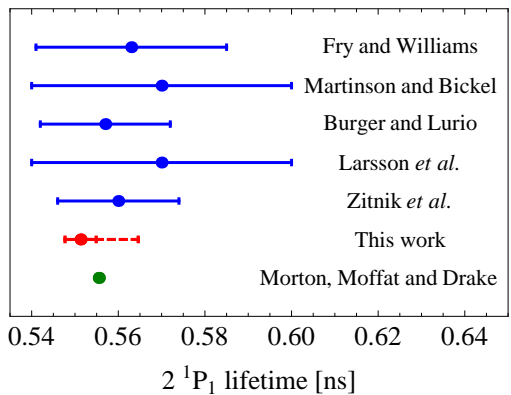}
		\caption{(Color online) Previously experimentally determined (blue) $2 \ ^1\text{P}_1$ lifetimes \cite{Fry1,*Martinson1,*Burger1,*Larsson1,*Zitnik1} compared to (red) our result and (green) the theoretical result by Morton, Moffat and Drake \cite{Morton2}. Our result contains an extended, dashed, error bar indicating a systematic uncertainty as discussed in the text.}
		\label{fig:Comparison_2_1P1_lifetimes}
	\end{center}
\end{figure}

Based on the same 77 linescans from which we determine the transition frequency, we find a natural linewidth of $289 \ (2)_{\text{stat}} \ (^{+ 0}_{- 7})_{\text{syst}}$~MHz. This corresponds to a lifetime of the $2 \ ^1\text{P}_1$ state of $0.551 \ (0.004)_{\text{stat}} \ (^{+ 0.013}_{- 0.000})_{\text{syst}}$~ns. This result is shown in Figure~\ref{fig:Comparison_2_1P1_lifetimes} together with previously determined lifetimes \cite{Fry1,*Martinson1,*Burger1,*Larsson1,*Zitnik1} and shows an improvement in the accuracy compared to the previous most accurate result. Our result is in agreement with the previous measurements, which are all based on completely different techniques, and agrees with a theoretical lifetime of 0.5555~ns, which is accurate to the last digit and calculated neglecting finite mass and relativistic effects that are expected below the 0.1\% accuracy level \cite{Morton2}.

To summarize, we have measured the $2 \ ^3\text{S}_1 \to 2 \ ^1\text{P}_1$ transition frequency in a quantum degenerate gas of $^4$He* to $1.6 \times 10^{-9}$ relative accuracy. From this measurement the $2 \ ^1\text{P}_1$ IE is determined with $6.7 \times 10^{-10}$ relative accuracy, in agreement with two recent independent determinations by Luo \emph{et al.} \cite{Luo1,*Luo1_erratum,Luo2}. We show a $>3\sigma$ discrepancy in the $2 \ ^1\text{P}_1$ IE with the most accurate QED calculation by Yerokhin and Pachucki \cite{Pachucki2}, indicating that a renewed effort on the QED calculations is required. We also report the most accurate determination of the $2 \ ^1\text{P}_1$ lifetime to date. This new determination is in agreement with theory and all previous experimental determinations.

We acknowledge the Foundation for Fundamental Research on Matter (FOM) for financial support through programme `Broken Mirrors and Drifting Constants'. We gratefully acknowledge K.S.E. Eikema for providing us the use of the frequency comb and its infrastructure. We also would like to thank J.S. Borbely, S. Knoop, J.C.J. Koelemeij, S.J.J.F.M. Kokkelmans and R.J. Rengelink for fruitful discussions and helpful suggestions.

\bibliography{High-precisionSpectroscopyInHelium_references}

\begin{thebibliography}{40}
\bibitem [{Foo()}]{Footnote1}%
  \BibitemOpen
  \href@noop {} {}\bibinfo {note} {{Here we ignore the alternative decay path
  $2 \ ^1\text{P}_1 \to 2 \ ^1\text{S}_0 \to 1 \ ^1\text{S}_0$, which also
  takes place but at a branching ratio of $0.1\%$ with the direct $2 \
  ^1\text{P}_1 \to 1 \ ^1\text{S}_0$ decay. Although the lifetime of the $2 \
  ^1\text{S}_0$ state is 50 ms, atoms in the $2 \ ^1\text{S}_0$ state feel an
  antitrapping potential in the ODT and the atoms will be pushed out of the
  trap and be observed as a loss, just like the atoms that would immediately
  decay to the $1 \ ^1\text{S}_0$ state.}}\BibitemShut {Stop}%
\end{thebibliography}
\bibliographystyle{apsrev4-1}

\onecolumngrid
\newpage

\section{Supplementary to `High-precision spectroscopy of the forbidden $2 \ ^3\text{S}_1 \to 2 \ ^1\text{P}_1$ transition in quantum degenerate metastable helium'}

In the main text we discuss several different calculations concerning the mean field shift, optical Bloch equations and an estimate of the nonlinear response of the MCP detector. Here we elaborate on these calculations.
\newline

\twocolumngrid

\section{Mean field shift}
The mean field shift, also known as the cold-collision shift, is caused by the fact that two $2 \ ^3\text{S}_1$ state atoms (which are our `initial state' atoms) have a different $s$-wave scattering length than the scattering length of a collision between a $2 \ ^3\text{S}_1$ state atom and a $2 \ ^1\text{P}_1$ state atom. This difference in scattering lengths leads to a different chemical potential for a gas of purely $2 \ ^3\text{S}_1$ atoms and a mixture of $2 \ ^3\text{S}_1$ and $2 \ ^1\text{P}_1$ atoms. It is this difference in chemical potential which forms an energy shift of the $2 \ ^3\text{S}_1$ and  $2 \ ^1\text{P}_1$ states that results in a shift of the resonance frequency.

The actual cold-collision cross-section $\sigma$ is not only a function of the scattering length $a$ between two atoms, but also of the collision energy that is described by the relative collision momentum $k$. The relative momentum $k$ is defined such that the total kinetic energy of the collision is $E_{kin} = \hbar^2 k^2 / 2 \mu$,  with $\mu$ the reduced mass of the system. To first order in $k$, the cross-section is
\begin{align}
\sigma = \frac{4 \pi a^2}{1 + k^2 a^2}.
\end{align}
In the limit of very low collision energy or scattering length, $ka \ll 1$, we obtain the hard-sphere scattering result $\sigma_0 = 4 \pi a^2$. In the other limiting case, $ka \gg 1$, the collision cross-section is bound as $\sigma (k) = 4 \pi / k^2$. This is known as the unitarity limit where the scattering length is so large that the collision cross-section only depends on the collision energy.

There is no \emph{a priori} indication of the sign and magnitude of the $2 \ ^3\text{S}_1$-$2 \ ^1\text{P}_1$ scattering length $a_{SP}$. Therefore we need a mean field shift model that is valid for both small values of $a_{SP}$ in the hard-sphere scattering regime and for very large $a_{SP}$ in the unitarity limit. A suitable model has been used by Kokkelmans \emph{et al.} to calculate the mean field shift and line broadening in a rubidium clock \cite{Kokkelmans1}. The mean field shift $\delta \omega$ and line broadening $\gamma$ of the $2 \ ^3\text{S}_1 \to 2 \ ^1\text{P}_1$ transition can be described by a sum over all relevant atomic states $j$ as
\begin{align}
i \delta \omega - \gamma = \sum_j{n_j \langle v (i \lambda_j - \sigma_j) \rangle},
\end{align}
with $\langle \rangle$ defining a thermal average, $n_j$ the density of atoms in state $j$, $v$ the relative velocity between two colliding particles and $\lambda_j$ and $\sigma_j$ the shift and width cross-sections. These cross-sections can be obtained from the S-matrices of the collisions as
\begin{eqnarray}
i \lambda_j - \sigma_j = (1 + \delta_{1j})(1 + \delta_{2j}) \frac{\pi}{k^2} \nonumber \\
\times \sum_l{(2l+1) [S^l_{(1j),(1j)}S^{l\star}_{(2j),(2j)}-1]},
\end{eqnarray}
with $l$ the partial wave index. In the $s$-wave collision regime we can limit ourselves to $l = 0$ and the S-matrices become
\begin{align}
S_{(nj),(nj)} = \frac{1 - i k a_{nj}}{1 + i k a_{nj}}.
\end{align}
For the $2 \ ^3\text{S}_1 \to 2 \ ^1\text{P}_1$ transition, the density of excited state atoms can be ignored due to the fast decay to the $1 \ ^1\text{S}_0$ ground state, and the summation over the states is reduced to a complex function $\Omega$ defined as
\begin{align}
\Omega \equiv i \delta \omega - \gamma = \frac{4 \pi \hbar}{m k} n_S \Big( \frac{1 - i k a_{SS}}{1 + i k a_{SS}} \cdot \frac{1 + i k a_{SP}}{1 - i k a_{SP}} -1 \Big).
\end{align}
Here $a_{SS}$ is the $2 \ ^3\text{S}_1$-$2 \ ^3\text{S}_1$ scattering length of the $^5\Sigma^+_g$ potential. From this final expression we can find the broadening and shift as $\gamma = -\text{Re}(\Omega)$ and $\delta \omega = \text{Im}(\Omega)$. 

The kinetic energy of the atoms in the BEC can be estimated from the width of the momentum distribution of the atoms in the BEC and we find an energy corresponding to a few nK. For the thermal atoms, the temperature is about 0.2 $\mu\text{K}$. The absorption of a 887~nm photon will increase the kinetic energy of a helium atom by about 3 $\mu\text{K}$ which is much larger than any of the kinetic energies of the thermal and condensed atoms. Therefore we ignore any initial velocity distribution of the gas and describe the collision energy solely by the recoil energy of the 887~nm photon. The peak density of the BEC (and therefore $n_S$) can be calculated using the chemical potential of the BEC extracted from the TOF momentum distribution fits as described in the main text. We find an average density of $2.0 \ (0.4) \times 10^{13} \ \text{cm}^{-3}$ and we use the scattering length $a_{SS} = 7.512(5)$~nm $= 142.0(1) \ a_0$, where $a_0$ is the Bohr radius \cite{Moal1}.

\begin{figure}[htbp]
	\begin{center}
	\includegraphics[width=0.4\textwidth]{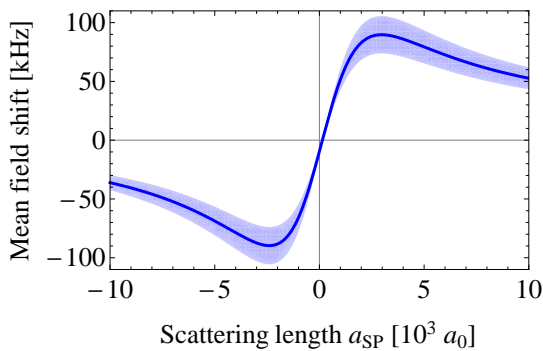}
	\caption{(Color online) Mean field shift calculated for any scattering length $a_{SP}$ in the range $-10^4 \ a_0 < a_{SP} < 10^4 \ a_0$. The shaded area is the uncertainty in the mean field shift due to the uncertainty in the average density as given in the text. The maximum shift is $\pm$~90~(20)~kHz at scattering lengths of $2700 \ a_0$ or $-2400 \ a_0.$}
	\label{fig:MFS}
	\end{center}
\end{figure}

\begin{figure}[htbp]
	\begin{center}
	\includegraphics[width=0.4\textwidth]{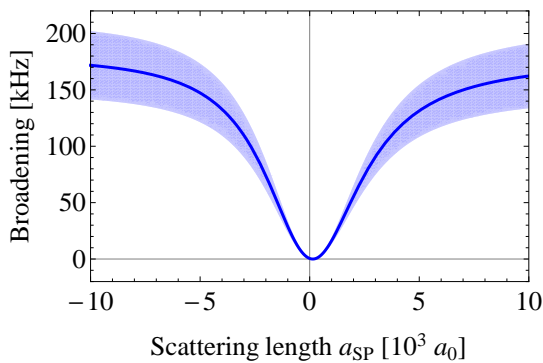}
	\caption{(Color online) Broadening calculated for any scattering length $a_{SP}$ in the range $-10^4 \ a_0 < a_{SP} < 10^4 \ a_0$, where the shaded area is the uncertainty in the broadening due to the uncertainty in the average density as given in the text.}
	\label{fig:MFS_broadening}
	\end{center}
\end{figure}

Using these parameters, the mean field shift and broadening can be calculated for any possible value of $a_{SP}$. The results are shown in Figures \ref{fig:MFS} and \ref{fig:MFS_broadening}. We find the maximum possible mean field shift to be $\pm$~90~(20)~kHz at scattering lengths $a_{SP} = 2700 \ a_0$ or $a_{SP} = -2400 \ a_0$. Furthermore, the maximum broadening is on the order to 100~kHz, which is negligible compared to the natural linewidth of 287~MHz. 

Although the shift represents almost 20\% of the statistical error bar on our determined transition frequency, the scattering lengths at which this would occur are very large. In order to `accidentally' have such a large scattering length, the collision energy should be resonant with a bound state in the  $2 \ ^3\text{S}_1$-$2 \ ^1\text{P}_1$ interatomic potential. The typical energy spacing of such bound states is on the order to 10's of GHz, and with a linewidth of approximately 300~MHz of the molecular bound state, the probability of being resonant is at the percent level and therefore quite unlikely. Furthermore, the used model does not incorporate the short lifetime of 0.555~ns of the $2 \ ^1\text{P}_1$ state. The finite lifetime would reduce the mean field shift even more, as it would suppress the interaction between the two atoms \cite{Julienne1}. These last two arguments allow us to assume that the mean field shift is much smaller than $\pm$~90~kHz and therefore negligible compared to the uncertainty in our determined transition frequency.

\section{Using the optical Bloch equations}
In our experiment we observe the transition by exciting $2 \ ^3\text{S}_1$ state atoms to the $2 \ ^1\text{P}_1$ state. As the lifetime of the excited state is only 0.555~ns and the measured transition is too weak for stimulated emission to occur, the excited atoms decay to the $1 \ ^1\text{S}_0$ ground state before they even have the chance to move  out of the trap potential (which is anti-trapping for the $2 \ ^1\text{P}_1$ state). The recoil kick of the emitted 58~nm photon is large enough for the ground state atom to leave the trap \cite{Footnote1}. Therefore any atom that is excited, will leave the trap. This means that even if the probe beam is off-resonant all atoms will eventually be lost from the trap. In the limiting case of infinitely long interaction time, one would then observe an infinitely broad transition. Obviously this is a broadening effect which should be corrected for. We calculate the population of the three states that are involved in this problem and evaluate the population of the initial state as a function of interaction time to see what happens to the linewidth as measured in the experiment.

\begin{figure}[tbp]
	\begin{center}
	\includegraphics[width=0.3\textwidth]{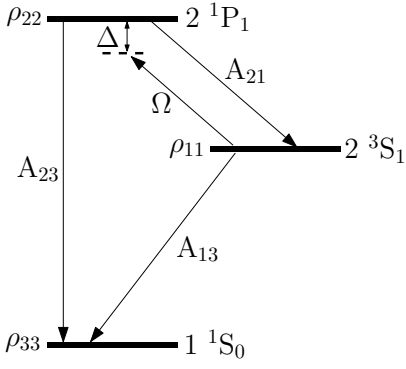}
	\caption{Schematic of the level scheme as used in the OBE model. The populations of the three atomic states are denoted $\rho_{11}, \rho_{22}$ and $\rho_{33}$, respectively. The different states are coupled through the decay channels $\text{A}_{23}$ and $\text{A}_{13}$ and the spontaneous emission rate $\text{A}_{21}$. The Rabi frequency $\Omega$ represents the interaction with the probe beam, which can have a detuning $\Delta$ with respect to the resonance frequency.}
	\label{fig:OBElevelscheme}
	\end{center}
\end{figure}

We use a simple three-level system consisting of the metastable $2 \ ^3\text{S}_1$ state, the excited $2 \ ^1\text{P}_1$ state and the $1 \ ^1\text{S}_0$ ground state \cite{Footnote1}. To verify the validity of this model, we have calculated the off-resonant scattering rates of the most probable transitions from the $2 \ ^3\text{S}_1$ and the $2 \ ^1\text{P}_1$ states to higher states either due to the 886.6~nm or 1557.3~nm light that is present during the measurement. The scattering rates are many orders of magnitude smaller than the pumping rate of the $2 \ ^3\text{S}_1 \to 2 \ ^1\text{P}_1$ transition, and photo-ionization of the $2 \ ^3\text{S}_1$ state due to reabsorption of a 58~nm photon emitted during the decay of the $2 \ ^1\text{P}_1$ state is also fully negligible. A schematic overview of the three-level system is shown in Figure \ref{fig:OBElevelscheme}. In this Figure we have added two decay channels $\text{A}_{13} = 1.272 \times 10^{-4} \ \text{s}^{-1}$, $\text{A}_{23} = 1.801 \times 10^9 \ \text{s}^{-1}$ and the Einstein A coefficient $\text{A}_{21} = 1.442 \ \text{s}^{-1}$ \cite{Drake4}.  The Rabi frequency in this system is defined via 
\begin{equation}
\Omega^2 = \frac{2 \pi c^2}{\hbar \omega_0^3} \text{A}_{21} I_0,
\end{equation}
where $I_0 = P/\pi w_0^2$ is the intensity of the spectroscopy beam, $P$ the spectroscopy beam power and $w_0$ the beam waist. The transition frequency is given by $\omega_0/2 \pi$. For a typical probe beam power $P \approx 50$~mW and a beam waist $w_0 \approx 2.5$~mm, the Rabi frequency $\Omega \approx 2\pi \times 7$~kHz, which is small as it is a forbidden transition.

\begin{figure}[tbp]
	\begin{center}
	\includegraphics[width=0.4\textwidth]{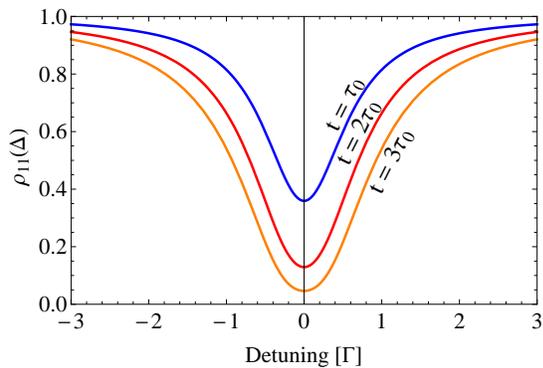}
	\caption{(Color online) Population of the initial $2 \ ^3\text{S}_1$ state ($\rho_{11}$) as function of the detuning $\Delta$ for different probe times $t$. As we increase the interaction time from $\tau_0 = 1$~s to $2 \tau_0$ and $3 \tau_0$, the lineshape becomes broader and deeper. This is the broadening effect due to the depletion of atoms in the trap.}
	\label{fig:OBE_examples}
	\end{center}
\end{figure}

The optical Bloch equations (OBEs) describing the populations of the states of this three-level system are based on the OBEs described by Van Leeuwen and Vassen for a similar system in helium \cite{Leeuwen1} and are
\begin{eqnarray}
\dot{\rho}_{11} &=& \text{A}_{21} \rho_{22} - \text{A}_{13} \rho_{11} + i \frac{\Omega}{2}(\rho_{21}-\rho_{12}), \nonumber \\
\dot{\rho}_{22} &=& - (\text{A}_{23} + \text{A}_{21})\rho_{22} - i \frac{\Omega}{2}(\rho_{21}-\rho_{12}), \\
\dot{\rho}_{33} &=& \text{A}_{23} \rho_{22} + \text{A}_{13} \rho_{11}. \nonumber
\end{eqnarray}
We can simplify this system of differential equations by writing $\text{A}_{13} = \Gamma_1$ and $\text{A}_{23} = \Gamma_2$, where $\Gamma_1/2\pi$ and $\Gamma_2/2\pi$ are the linewidths of the $2 \ ^3\text{S}_1$ and $2 \ ^1\text{P}_1$ states, respectively. Applying the approximation $\Gamma_2 \gg \text{A}_{21} \gg \Gamma_1$, the OBEs describing the coherences between the states are
\begin{eqnarray}
\dot{\rho}_{12} &=& - \Big( \frac{\Gamma_1 + \Gamma_2}{2} + i\Delta \Big) \rho_{12} + i \frac{\Omega}{2}(\rho_{22}-\rho_{11}), \nonumber \\
\dot{\rho}_{13} &=& - \frac{\Gamma_1}{2} \rho_{13}, \\
\dot{\rho}_{23} &=& - \frac{\Gamma_2}{2} \rho_{23}, \nonumber
\end{eqnarray}
where $\rho_{ij} = \rho^{\dagger}_{ji}$. Introducing the parameters $p~=~(\rho_{12}~+~\rho_{21})/2$ and $q~=~i(\rho_{21}~-~\rho_{12})/2$, we get
\begin{eqnarray}
\dot{\rho}_{11} &=& \Omega q, \nonumber \\
\dot{\rho}_{22} &=& - \Gamma_2 \rho_{22} - \Omega q,\nonumber  \\
\dot{\rho}_{33} &=& \Gamma_2\rho_{22}, \nonumber \\
\dot{p} &=& - \frac{\Gamma_2}{2} p + \Delta q, \\
\dot{q} &=& - \frac{\Gamma_2}{2}q - \Delta p + \frac{\Omega}{2}(\rho_{22}-\rho_{11}), \nonumber \\
\dot{\rho}_{13} &=& 0. \nonumber 
\end{eqnarray}
We can solve this system of differential equations analytically using a symbolic mathematics solver such as Mathematica. The full solution of the population of the initial state, $\rho_{11}$, follows from
\begin{eqnarray}
\text{e}^{t\frac{\Gamma_2}{2}} \rho_{11}(t) = \frac{\sqrt{S} + \Gamma_2^2 + 4 \Delta^2}{2 \sqrt{2}} \text{cosh} \Bigg( \frac{t}{\tau_1} \Bigg) \nonumber \\
+ \frac{\Gamma_2}{\sqrt{2}} \frac{\sqrt{S}+D^+}{\sqrt{S}\sqrt{\sqrt{S} + D^-}} \text{sinh} \Bigg( \frac{t}{\tau_1} \Bigg) \nonumber \\
+ \frac{\sqrt{S} - \Gamma_2^2 - 4 \Delta^2}{2 \sqrt{2}} \text{cos} \Bigg( \frac{t}{\tau_2} \Bigg) \nonumber \\
+ \frac{\Gamma_2}{\sqrt{2}} \frac{\sqrt{S}-D^+}{\sqrt{S}\sqrt{\sqrt{S} - D^-}} \text{sin} \Bigg( \frac{t}{\tau_2} \Bigg)
\end{eqnarray}
where the parameters $S$,$D^{\pm}$, $\tau_1$ and $\tau_2$ are defined as
\begin{eqnarray}
S &\equiv& (4\Delta^2 + (\Gamma_2-2\Omega)^2) (4\Delta^2 + (\Gamma_2+2\Omega)^2), \\
D^{\pm} &\equiv& \Gamma_2^2 - 4\Omega^2 \pm 4\Delta^2, \\
\tau_1 &\equiv& \frac{2 \sqrt{2}}{\sqrt{\sqrt{S} + D^-}}, \\
\tau_2 &\equiv& \frac{2 \sqrt{2}}{\sqrt{\sqrt{S} - D^-}}.
\end{eqnarray}
In these equations we recognize the trigonometric terms that represent the coherent excitation, but in our parameter regime these do not play a significant role. The hyperbolic terms are related to the decoherence caused by the decay channel and represent the loss of atoms from the ODT. If we fill in the experimental parameters and look at the population of the initial state as a function of the detuning of the probe light, we obtain the lineshape as it should be observed in the experiment. This is shown in Figure \ref{fig:OBE_examples} for three different interaction times. It is clear that the line becomes broader and deeper as the interaction time increases, which is caused by the loss of atoms from the trap to the $1 \ ^1\text{S}_0$ state.

\begin{figure}[tbp]
	\begin{center}
	\includegraphics[width=0.4\textwidth]{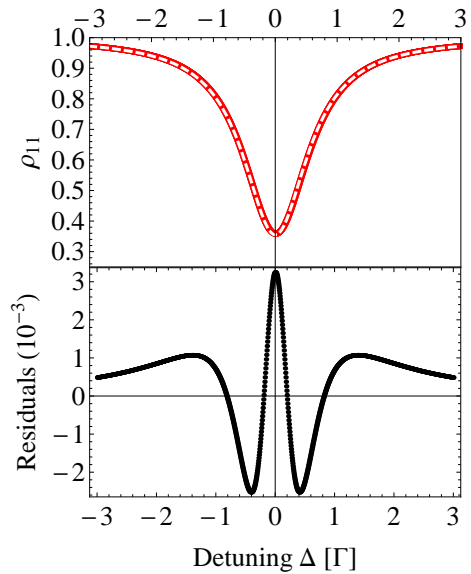}
	\caption{(Color online) Lorentzian fit (white dashed line) to the analytical model (red line) and the residuals (black dots). From the residuals, which are at the $10^{-3}$ level, we infer that the Lorentzian fit is good to the percent level, but is `quenched' compared to the analytical model. This means that the magnitude of the central peak is larger in the fit than it actually is, and the linewidth of the Lorentzian fit is slightly smaller than the actual linewidth of the curve.}
	\label{fig:OBE_LorFit}
	\end{center}
\end{figure}

The function $\rho_{11}(\Delta)$ as shown in Figure \ref{fig:OBE_examples} is a function dependent on many parameters and in our experiment we use a simple Lorentzian lineshape function defined as
\begin{eqnarray}
y(x) = C_0 - C_1 \Big(\frac{C_2}{2} \Big)^2 \frac{1}{(x-C_3)^2 + (C_2/2)^2},
\end{eqnarray}
where $C_0$ represents the background, $C_1$ the amplitude of the Lorentzian (which we also call the `depletion' and is related to $\rho_{11}$ as $C_1 = 1 - \rho_{11} (\Delta = 0)$), $C_2$ is the FWHM and $C_3$ the center frequency. As $\rho_{11}(\Delta)$ and a Lorentzian function are both symmetric functions, this has no effect on the determination of the transition frequency. To extract the linewidth from these fits we compare the analytical result $\rho_{11}(\Delta)$ with a Lorentzian lineshape and correct for any deviations between these two functions. The lifetime of the $2 \ ^1\text{P}_1$ state leads to a Lorentzian distribution in the frequency domain. The fact that we have a slightly different lineshape in our experiment is caused by the way the measurement is performed. However, by directly correcting the fitted Lorentzian distribution for deviations from the analytical lineshape, the lifetime of the  $2 \ ^1\text{P}_1$ state can still be extracted from these measurements. An example of a Lorentzian fit to the analytical result for typical experimental conditions is shown in Figure \ref{fig:OBE_LorFit}. From the structure in the residuals, we conclude that the Lorentzian is accurate beyond the 1\% level, after which it deviates from the analytical model most at the center and around $\Delta = \pm \Gamma_0/2$. This problem is parametrized by calculating the difference between the Lorentzian FWHM and the analytical model FWHM as a function of the on-resonance depletion of the trap (simply called `depletion'). As the depletion gets larger, the discrepancy between the Lorentzian function and the analytical model increases. For a depletion in the range of 0.4-0.6 (i.e. 40\%-60\% on-resonance loss of atoms, which corresponds to a $\rho_{11} (\Delta = 0)$ of 0.6-0.4), the correction factor on the FWHM is on average 0.6\% and relevant at our level of accuracy. The correction on the depletion is an order of magnitude smaller.

\begin{figure}[tbp]
	\begin{center}
	\includegraphics[width=0.4\textwidth]{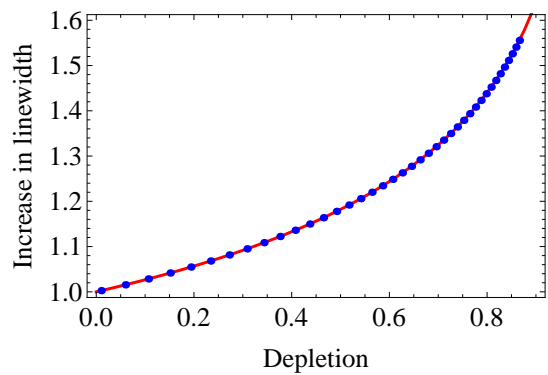}
	\caption{(Color online) Relative increase of the Lorentzian linewidth compared to the natural linewidth as function of the on-resonance depletion of the trap. The blue points are calculated from Lorentzian fits and the red curve is an spline-based interpolation function.}
	\label{fig:OBE_FWHM_increase}
	\end{center}
\end{figure}

The previous two corrections are related to the difference between the Lorentzian lineshape and the analytical model. The next issue, and the reason why we need to calculate the OBEs, is the linewidth broadening due to depletion of the trap. By comparing the linewidth determined by a Lorentzian fit to the linewidth used in the analytical model, the relative increase in linewidth as function of depletion of the trap can be calculated. The result is shown in Figure \ref{fig:OBE_FWHM_increase}. Although the increase in linewidth can be calculated to arbitrary precision, the deviation between the Lorentzian lineshape and the analytical model becomes larger than our accuracies for a depletion $> 0.75$ and the comparison of both models becomes less reliable. However, in our experiments our largest depletion is about $0.6$ and this effect is still smaller than our corrections. In our measurements the linewidth is increased on average by 23\% just because of depletion of the trap. As we determine the trap depletion experimentally, we can calculate the correction for the linewidth as measured in the experiment and obtain the natural linewidth. Any experimental uncertainties in the depletion and the linewidth are propagated through all aforementioned corrections. 

\section{Saturation effects of the MCP detector}
In the main text we mention possible saturation effects of the MCP detector that could also lead to line broadening. We expect saturation of the detector based on a simple estimation of the atom flux on the MCP and the typical dead time of the single channels. This calculation assumes a homogeneous distribution of the atoms arriving at the MCP detector both in space and time. Although it is an overestimate, it is further assumed that a single channel has zero response during its dead time. Assuming an atom flux of approximately $10^8$ atoms per second (all atoms arrive within 20~ms) and $10^6$ channels with a dead time of approximately $1$~ms, approximately 25\% of the channels will be saturated as the atoms hit the detector. We therefore cannot ignore the possibility of saturation effects, which have been observed in other BEC experiments with metastable helium \cite{Schellekens1}.

\begin{figure}[tbp]
	\begin{center}
	\includegraphics[width=0.4\textwidth]{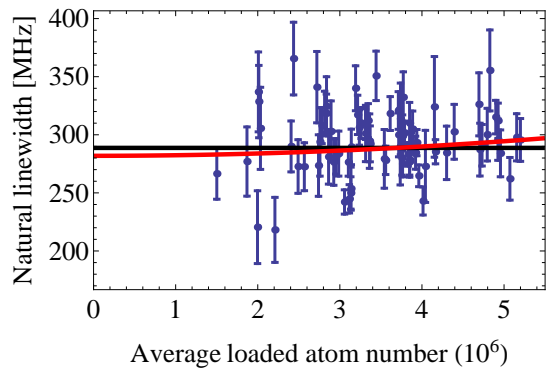}
	\caption{(Color online) Experimentally determined linewidths as a function of the total number of atoms loaded into the ODT. The horizontal black line represents the statistical average of 289~(2)~MHz. The red line is a quadratic function as mentioned in the text, which gives a natural linewidth of 282~(5)~MHz and a quadratic term coefficient of $C_2 = 5 \ (6) \times 10^{-13}$~MHz/atom.}
	\label{fig:MCP_nonlinearity}
	\end{center}
\end{figure}

It is not possible to calibrate an MCP detector to sub-1\% accuracy, so we analyze our data using a simple model to search for any possible nonlinearity. Our experimentally determined natural linewidths of the transition are shown in Figure \ref{fig:MCP_nonlinearity} as a function of the number of atoms that are loaded into the ODT during the measurement. We show the average and a quadratic dependance $y(x) = C_0 + C_2 x^2$. The resulting coefficients of the fits are $C_0 = 282 \ (5)$ MHz and $C_2 = 5 \ (6) \times 10^{-13}$~MHz/atom. The nonlinear parameter has a value that is different from zero only at the $1\sigma$ level and therefore we can not conclude that there is a clear nonlinear effect. However, the determined natural linewidth shifts to a slightly lower value, which is indeed the broadening effect expected for saturation of the MCP detector. Therefore we add a systematic uncertainty of -7~MHz to our final result which indicates the possible shift if we allow a nonlinear response of the MCP detector.

\end{document}